\documentstyle[epsfig,amssymb,floatflt,float,here]{mn2e}

\newif\ifAMStwofonts

\newcommand{\kms}{km~s$^{-1}$\,}
\newcommand{\D}{\displaystyle}

\newcommand {\hi} {H\,{\small I}\,}
\newcommand {\sm}{$M_{\odot}$\,}

\title[The radial velocity dispersion of profile]{The radial velocity
dispersion profile of the Galactic halo: Constraining the density
profile of the dark halo of the Milky Way} \author[G. Battaglia et
al.]  {Giuseppina Battaglia$^1$\thanks{Corresponding author. E-mail:
gbattagl@astro.rug.nl}, Amina Helmi$^1$, Heather Morrison$^2$, Paul
Harding$^2$, \newauthor Edward W. Olszewski$^3$, Mario Mateo$^4$,
Kenneth C. Freeman$^5$, John Norris$^5$, and \newauthor Stephen
A. Shectman$^6$ \\ \\ $^1$Kapteyn Astronomical Institute, University
of Groningen, P.O.Box 800, 9700 AV Groningen, The Netherlands\\
$^2$Astronomy Department, Case Western Reserve University, Cleveland,
OH 44106\\ $^3$Steward Observatory, University of Arizona, Tucson, AZ
85721\\ $^4$Astronomy Department, University of Michigan, Ann Arbor,
MI48109\\ $^5$Research School of Astronomy \& Astrophysics, The
Australian National University, Mount Stromlo Observatory, Cotter
Road, \\ Weston ACT 2611\\ $^6$Carnegie Observatories, 813 Santa
Barbara Street, Pasadena, CA 91101 }

\pagerange{\pageref{firstpage}--\pageref{lastpage}}
\pubyear{}

\begin{document}

\maketitle

\label{firstpage}

\begin{abstract}

We have compiled a new sample of 240 halo objects with accurate
distance and radial velocity measurements, including globular
clusters, satellite galaxies, field blue horizontal branch stars and
red giant stars from the Spaghetti survey. The new data lead to a
significant increase in the number of known objects for Galactocentric
radii beyond 50 kpc, which allows a reliable determination of the
radial velocity dispersion profile out to very large distances. The
radial velocity dispersion shows an almost constant value of 120 \kms
out to 30 kpc and then continuously declines down to 50 \kms at about
120 kpc.  This fall-off puts important constraints on the density
profile and total mass of the dark matter halo of the Milky Way. For a
constant velocity anisotropy, the isothermal profile is ruled out,
while both a dark halo following a truncated flat model of mass
$1.2^{+1.8}_{-0.5}\times 10^{12}$\sm and an NFW profile of mass
$0.8^{+1.2}_{-0.2}\times 10^{12}$\sm and $c=$18 are consistent with
the data.  The significant increase in the number of tracers combined
with the large extent of the region probed by these has allowed a more
precise determination of the Milky Way mass in comparison to previous
works. We also show how different assumptions for the velocity
anisotropy affect the performance of the mass models.
\end{abstract}


\section{Introduction}

The determination of the total mass of the Galaxy has been a subject
of considerable interest since the work of Kapteyn in the early 1920s
(see Fich \& Tremaine 1991 for a nice introductory review on the
subject). Since then, the mass of the Milky Way has seen its estimates
grow by factors of ten to a hundred, with some dependence on the type
of mass tracer used: \hi kinematics, satellite galaxies and globular
clusters, or the Local Group infall pattern. The most recent
determinations yield fairly consistent values for the mass within 50
kpc, with an uncertainty of the order of 20\% for a given mass model 
(Kochanek 1996;
Wilkinson \& Evans 1999, hereafter W\&E99; 
Sakamoto, Chiba \& Beers 2003, hereafter SCB03). However, even
today, the total mass of the Galaxy is not known better than within a
factor of two.

Whatever method is used, be it the \hi kinematics, globular clusters,
satellite galaxies, or halo giants, it is only possible to determine
the mass enclosed in the region probed by these tracers (Binney \&
Tremaine 1987). This implies that the rotation curve derived from \hi
will only constrain the mass within roughly 18 kpc from the Galactic
centre (Rohlfs \& Kreitschmann 1988; Honma \& Sofue 1997), a region
which is baryon dominated. Globular clusters and satellite galaxies
are, in principle, better probes of the large scale mass distribution
of the Galaxy, since they are found out to distances beyond 100
kpc. However, there are only 15 such objects beyond 50 kpc (Zaritsky
et al. 1989; Kochanek 1996). Only 6 of these have proper motion
measurements, which despite the large errors, can further constrain the
shape of the velocity ellipsoid. Using this dataset, W\&E99
favour isotropic to slightly tangentially anisotropic models,
although 1$\sigma$ contours for the velocity anisotropy $\beta$ 
give $-0.4 \lesssim \beta \lesssim 0.7$. 
SCB03 have added to the sample used by W\&E, 
field blue
horizontal branch stars with proper motions and radial
velocities. While this is clearly an improvement, these stars are
located within 10 kpc of the Sun, which strongly limits their constraining
power at larger radii. In their models, the velocity ellipsoid is
tangentially anisotropic, with $\beta \sim -1.25$ as the most
likely value. 

It is clearly important to measure the total mass of the Galaxy in
order to constrain its dark-matter content. However, it is also
critical to determine its distribution: density profile, flattening,
velocity ellipsoid, etc.  One of the most fundamental predictions of
cold-dark matter models is that the density should follow an NFW
profile throughout most of the halo (Navarro, Frenk \& White
1997). 

The density profiles derived from the gas rotation curves of
large samples of external galaxies do not always follow the NFW shape
(de Blok et al. 2001). Tracers at larger distances are rare, but
objects such as planetary nebulae or globular clusters could yield
powerful constrains on the mass distribution at those radii, for example 
for elliptical galaxies as shown by Romanowsky et al. (2003).

In the case of the Milky Way, the situation is not dissimilar. The
distribution of mass inside the Solar circle has been studied
extensively (see e.g. Dehnen \& Binney 1998; Evans \& Binney 2001;
Bissantz, Debattista \& Gerhard 2004). A common conclusion is that
there is little room for dark-matter in this region of the Galaxy. 

But does the dark-matter beyond the edge of the Galactic disk follow
an NFW profile? How does the most often assumed isothermal profile
perform in this region of the Galaxy (e.g. Sommer-Larsen et al. 1997;
Bellazzini 2004)? Is the velocity ellipsoid close to isotropic as
found in CDM simulations (Ghigna et al. 1998)?  Modeling of the
kinematics of halo stars by Sommer-Larsen
et al. (1997) favoured an ellipsoid that became more tangentially
anisotropic towards larger distances, while Ratnatunga \& Freeman
(1989) found a constant line-of-sight velocity dispersion out to 25
kpc. 

These fundamental issues can only be addressed when a sufficiently
large number of probes of the outer halo of the Galaxy are
available. Ideal tracers are red giant stars or blue horizontal branch
stars, which can be identified photometrically also at large
galactocentric distances (Morrison et al. 2000; Clewley et al. 2002;
Sirko et al. 2004a,b). Spectroscopic follow-up allows both the
confirmation of the luminosity class as well as the determination of
radial velocities with relatively small errors (Morrison et al. 2003).
With the advent of wide field surveys, such as the Sloan Digital Sky
Survey, or the Spaghetti survey, the numbers of such outer
halo probes have increased by large amounts, making this an ideal time
to address the mass distribution of our Galaxy in greater detail.

This paper is organized as follows. In the next section we describe
the observational datasets used to determine the radial velocity
dispersion curve. In Sec. \ref{sec:models} we introduce several mass
models for the dark halo of our Galaxy and derive how the line of
sight velocity dispersion depends on the model parameters. In
Sec. \ref{sec:results} we compare the data to the models and derive the
best fit values of the parameters using $\chi^2$ fitting. Finally we
discuss our results and future prospects in Sec. \ref{sec:disc}.
%
\section{The radial velocity dispersion curve}

\subsection{The observational datasets}

\begin{table} 
\begin{center}
\begin{tabular}{lcc}
\hline
Objects           & number of objects  &  Source           \\
\hline
Globular clusters  & 44    &   Harris (1997)   \\
FHB stars          & 130    &   Wilhelm et al. (1999b), \\
      & &                         Clewley et al. (2004) \\
Red halo giants    & 57    &    Spaghetti survey      \\
Satellite galaxies & 9     &    Mateo (1998)           \\
\hline
\end{tabular}
\caption{Characteristics of the data used in this paper. In all cases,
position in the sky, heliocentric distance and line of sight
velocities are available. 
} \label{tab:tab1}
\end{center}
\end{table}

Our goal is to derive the radial velocity dispersion profile of the
Milky Way stellar halo in the regime where it is dominated by the
gravitational potential of its dark-matter halo. Hence we restrict
ourselves to tracers located at Galactocentric distances greater than
10 kpc, where the disc's contribution is less important.

\begin{figure}
\begin{center}
\includegraphics[width=85mm]{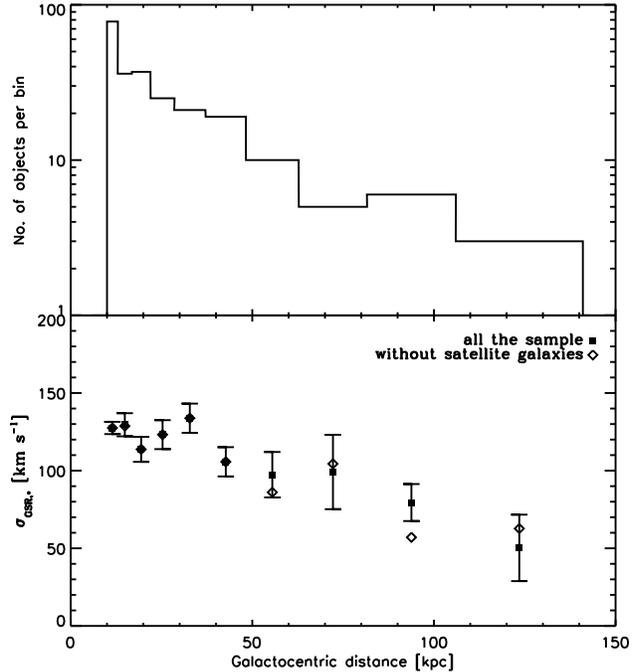}
\caption{The top panel shows the number of objects per bin  
in our sample. The bottom panel shows the
Galactocentric radial velocity dispersion of the Milky Way halo. 
The squares with error-bars
correspond to the dispersion profile for the whole sample.
The diamonds indicate the Galactocentric radial velocity dispersion if the
satellite galaxies are not included in the sample.}
\label{fig:disp_data}
\end{center}
\end{figure}

We use a sample of 9 satellite galaxies, 44 globular clusters, 57 halo
giants and 130 field blue horizontal branch stars (FHB).  The various
data sources of this sample are listed in Table \ref{tab:tab1}. It is
worth noting that there are 24 objects located beyond 50 kpc in our
sample, and that we have enough statistics to measure radial velocity
dispersion out to 120 kpc as shown in the top panel of
Figure~\ref{fig:disp_data}. This covers a significantly larger radial
range than many previous works, including e.g. Sommer-Larsen et
al. (1997), whose outermost point is at 50 kpc.

The red halo giants are from the ``Spaghetti'' Survey (Morrison et
al. 2000). This is a pencil beam survey that has so far covered 20
$\deg^2$ in the sky, down to $V \sim 20$. It identifies candidate halo
giants using Washington photometry, where the 51 
filter\footnote{The 51 filter is centered on
the Mgb/MgH feature near 5170 $\dot{\rm{A}}$} allows for a
first luminosity selection. Spectroscopic observations are then
carried out to confirm the photometric identification and to determine
the radial velocities of the stars.

We have derived the heliocentric distance for the FHB stars 
from Wilhelm et al. (1999b) using the relation 
$$ M_V({\rm HB})= 0.63 + 0.18 ({\rm [Fe/H]} + 1.5)$$ (Carretta et
al. 2000).  

In all cases, accurate distances and radial velocities are
available: the average error in velocity ranges from a few \kms
(satellite galaxies and globular clusters) to 10-15 \kms (FHB stars
and red giants); the typical relative distance error is approximately
10\%.

\begin{figure}
\begin{center}
\includegraphics[width=85mm]{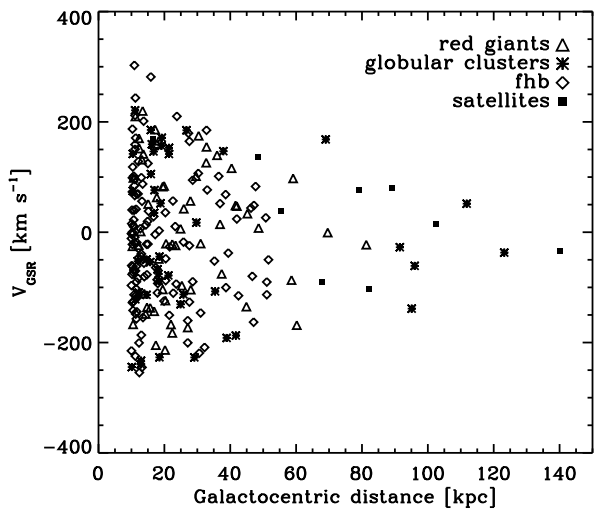}
\caption{Heliocentric l.o.s. velocities corrected for the Solar Motion
and the LSR motion ($V_{\rm GSR}$) for the sample used in this work
(triangles: red giants; asterisks: globular clusters; diamonds: field
horizontal branch stars; filled squares: satellite galaxies).}
\label{fig:vhel}
\end{center}
\end{figure}

When transforming the heliocentric l.o.s. velocities, $V_{\rm los}$,
into Galactocentric ones, $V_{\rm GSR}$, we assume a circular velocity
of $V_{\rm LSR} = 220$~km s$^{-1}$ at the solar radius ($R_{\odot} = $
8 kpc) and a solar motion of ($U$,$V$,$W$) $=$ (10, 5.25, 7.17) \kms,
where $U$ is radially inward, $V$ positive in the direction of the
Galactic rotation and $W$ towards the North Galactic Pole (Dehnen \&
Binney 1998).  Hereafter we refer to: the radial velocity (dispersion)
measured in a heliocentric coordinate system as the l.o.s. velocity,
$V_{\rm los}$ (dispersion, $\sigma_{\rm los}$); the l.o.s. velocity
(and its dispersion) corrected for the solar motion and the LSR motion
as the Galactocentric radial velocity, $V_{\rm GSR}$ (dispersion,
$\sigma_{\rm GSR}$); the radial velocity (and its dispersion) in a
reference frame centered on the Galactic Centre as the true radial
velocity, $V_r$ (dispersion, $\sigma_r$).  Figure \ref{fig:vhel} shows
$V_{\rm GSR}$ as function of the Galactocentric distance $r$ for all
the objects used in this work.

The bottom panel in Figure \ref{fig:disp_data} shows the
Galactocentric radial velocity dispersion as function of distance from
the Galactic centre. This is computed in bins whose width is
approximately twice the average distance error of objects in the
bin. This implies that our bin sizes range from 3 kpc at $r \sim 10$
kpc, to 40 kpc at $r \sim 120$ kpc.  The error-bar on the velocity
dispersion in each bin is calculated performing Monte Carlo
simulations. We assume the velocity and distance errors are gaussianly
distributed in the heliocentric reference frame. In practice, this 
means that we randomly generate velocities and distances for each one
of the stars, whose mean and dispersion are given by the observed
value and its estimated error, respectively.  We then convert the
heliocentric quantities into Galactocentric ones.  We repeat this
exercise for 10,000 sets, and for each of these we measure $\sigma_{\rm
GSR}$ in the same bins as the original data. We use the rms of this
velocity dispersion, obtained from the 10,000 simulations, as the error
on the velocity dispersion we measured in the bin.

One may question whether the satellite galaxies can be considered fair
tracers of the gravitational potential of the dark matter halo of the
Milky Way (e.g. Taylor, Babul \& Silk 2004; Gao et al. 2004). To get a
handle on this issue, we compute the velocity dispersion profile both
with and without them (squares and diamonds, respectively in
Fig. \ref{fig:disp_data}). Since the trend is similar in both cases we
may consider the satellites to be reliable probes of the outer halo
potential.

\subsection{The models}
\label{sec:models}
\subsubsection{Jeans equations}

If we assume that the Galactic halo is stationary and spherically
symmetric we can derive the (expected) radial velocity dispersion
profile $\sigma_{r,*}$ of the stars from the Jeans equation (Binney \&
Tremaine 1987):
\begin{equation}
\frac{1}{\rho_*}\frac{d({\rho_* \sigma_{r,*}^2})}{dr} + \frac{2\beta
\sigma_{r,*}^2}{r} = -\frac{d\phi}{dr} = - \frac{V_{\rm c}^2}{r}
\label{eq:jeans}
\end{equation}
where $\rho_*(r)$ is the mass density of the stellar halo, $\phi(r)$
and $V_{\rm c}(r)$ are the potential and circular velocity of the dark
matter halo and $\beta$ is the velocity anisotropy parameter, defined
as $\D \beta = 1 - \frac{\sigma_{\theta}^2}{\sigma_r^2}$, and assuming
$\sigma_{\theta}^2=\sigma_{\phi}^2$. Note that $\beta = 0$ if the
velocity ellipsoid is isotropic, $\beta = 1$ if the
ellipsoid is completely aligned with the radial direction, while
$\beta < 0$ for tangentially anisotropic ellipsoids.  

The Jeans equation allows us to determine a unique solution for the
mass profile if we know $\sigma_{r,*}^2(r)$, $\rho_*(r)$ and
$\beta(r)$, although this solution is not guaranteed to produce a
phase-space distribution function that is positive everywhere. We are,
however, faced with two uncertainties: the velocity anisotropy and the
behaviour of the stellar halo density at very large distances.  The
latter has been determined to vary as a power-law $\rho_*(r) \propto
r^{-\gamma}$ with $\gamma\sim$ 3.5 out to $\sim 50$ kpc (Morrison et
al. 2000; Yanny et al. 2000), and we shall assume this behaviour can
be extrapolated all the way out to our last measured point. More
crucial is the unknown variation of the velocity anisotropy with
radius, which is difficult to determine because of the lack of tracers
with accurate proper motions beyond the Solar neighbourhood. This
implies in principle, that large amounts of kinetic energy can be
hidden to the observer, an effect known as the mass-velocity
anisotropy degeneracy.  For sake of simplicity, and given that the
situation is unlikely to change until the advent of new space
astrometric missions such as SIM and Gaia (Perryman et al. 2001),
throughout most of this work we shall make the assumption that $\beta$
is constant, i.e. independent of radius $r$.

To derive Eq.~(\ref{eq:jeans}) we have assumed that the stellar halo
can be considered as a tracer population of objects moving in an
underlying potential. This is justified by the negligible amount of
mass present in this component, compared to, for example, that in the
disk and the dark halo.

The (expected) radial velocity dispersion for the tracer population $
\sigma_{r,*}$ may be thus derived by integrating
Eq.~(\ref{eq:jeans}). This leads to 
\begin{equation}
\sigma_{r,*}^2(r) = {1\over{\rho_*\,e^{\int 2\beta dx}}}{\int_x}^{\infty}
\rho_*\,V_{\rm c}^2 \,e^{\int 2\beta dx''}\,dx', \qquad x= \ln r.
\label{eq:sigma}
\end{equation}
Here, we have used that $r^{2 \beta} \rho_* \sigma_{r,*}|_\infty = 0$.
Note that the radial velocity dispersion of the tracer population
depends on the particular form of the circular velocity of the
underlying (gravitationally dominant) mass distribution.
 
Since proper motions are not available for the whole sample and we
only have access to heliocentric velocities, the quantity that we
measure is not the true radial velocity dispersion but
$\sigma{\rm_{GSR,*}}$.  When comparing this quantity to model
predictions, we must take in account a correction factor for the lack
of information on the tangential component of the velocity. Following
the procedure described in Appendix A, we find that the Galactocentric
radial velocity dispersion, $\sigma{\rm_{GSR,*}}$, is related to the
true radial velocity dispersion, $\sigma_{r,*}$ as
\begin{equation}
{\sigma{\rm_{GSR,*}}}(r) = {\sigma_{r,*}}(r)\,\sqrt{1 + 2\,
( 1- \beta) H(r)}, 
\end{equation}
where 
\begin{equation}
H(r) = \frac{r^2 + R_{\odot}^2}{4r^2} - 
\frac{{(r^2- R_{\odot}^2)}^2}{8r^3R_{\odot}}{\rm ln}\frac{r+R_{\odot}}
{r-R_{\odot}}.
\end{equation}
The above equation for $H(r)$ is valid at Galactocentric distances
$r>R_{\odot}$. For a purely radial anisotropic ellipsoid
($\beta=$ 1) $\sigma{\rm_{GSR,*}}$ and $\sigma_{r,*}$ coincide. For a
tangentially anisotropic stellar halo, the correction factor becomes
negligible at distances larger than about 30-40 kpc.  

\subsubsection{Specifing dark-matter halo models}

We adopt three different models for the spherically symmetric 
dark-matter halo potential:
\begin{itemize}
\item {\it Pseudo-Isothermal sphere.} This model has been extensively
used in the context of extragalactic rotation curve work.  The density
profile and circular velocity associated to a pseudo-isothermal sphere
are:
\begin{equation}
 \rho(r) = \rho_0\frac{r_{\rm c}^2}{(r_{\rm c}^2 + r^2)},
\end{equation}
and 
\begin{equation}
V_{\rm c}^2(r) = V_{\rm c}^2(\infty) \bigl(1 - \frac{r_{\rm c}}{r} {\rm
arctg}\frac{r}{r_{\rm c}}\bigr),
\end{equation}
where $r_{\rm c}$ is the core radius, and $\D \rho_0 = \frac{V_{\rm
c}^2(\infty)}{4 \pi G r_{\rm c}^2}$.  We set $V_{\rm c}(\infty)=$ 220
\kms as asymptotic value of the circular velocity.  At large radii the
density behaves as $\rho\propto r^{-2}$ giving a mass that increases
linearly with radius.
\\

\item {\it NFW model.} In this case the dark matter density profile is
given by
\begin{equation}
\rho(r) = \frac{\delta_c \rho_{\rm c}^0}{(r/r_{\rm s})(1+ r/r_{\rm s})^2}
\end{equation}
where $r_{\rm s}$ is a scale radius, $\rho_{\rm c}^0$ the present
critical density and $\delta_{\rm c}$ a characteristic overdensity.
The latter is defined by $\D \delta_{\rm c}= \frac{100\, c^3g(c)}{3}$,
where $c = r_{\rm v}/r_{\rm s}$ is the concentration parameter of the
halo and $ \D g(c) = \frac{1}{{\ln}(1+c) - c/(1+c)} $.  The circular
velocity associated with this density distribution is
\begin{equation}
V_{\rm c}^2(s) = {{V_{\rm v}^2
g(c)}\over{s}}\biggl[\,\ln(1+cs)-{cs\over{1+cs}}\biggr]
\end{equation}
where $V_{\rm v}$ is the circular velocity at the virial radius
$r_{\rm v}$ and $s=r/r_{\rm v}$. The concentration $c$ has been found
to correlate with the virial mass of the halo (Navarro, Frenk \& White
1997; Bullock et al. 2001; Wechsler et al. 2002).  However, the
relation presents a large scatter.  For example, for a halo of mass
1.0$\times 10^{12} h^{-1} M_{\odot}$ the predicted concentration
ranges between 10 and 20.  Hence, we cannot consider the NFW density
profile as a one-parameter family; we need to describe it by the
concentration $c$, and by the virial mass or the circular velocity at
the virial radius.  At large radii (for $r \gg r_{\rm s}$), the
density behaves as $\rho\propto r^{-3}$, and therefore, the total mass
diverges logarithmically.  However, we can impose that the particles
must be bound at the virial radius, and so when integrating Eq.~(2),
we set the upper integration limit to $r_{\rm v}$ and we use $r^{2
\beta} \rho_* \sigma_{r,*}|_{r_{\rm v}} = 0$.  \\
\begin{figure*}
\begin{center}
\includegraphics[width=85mm]{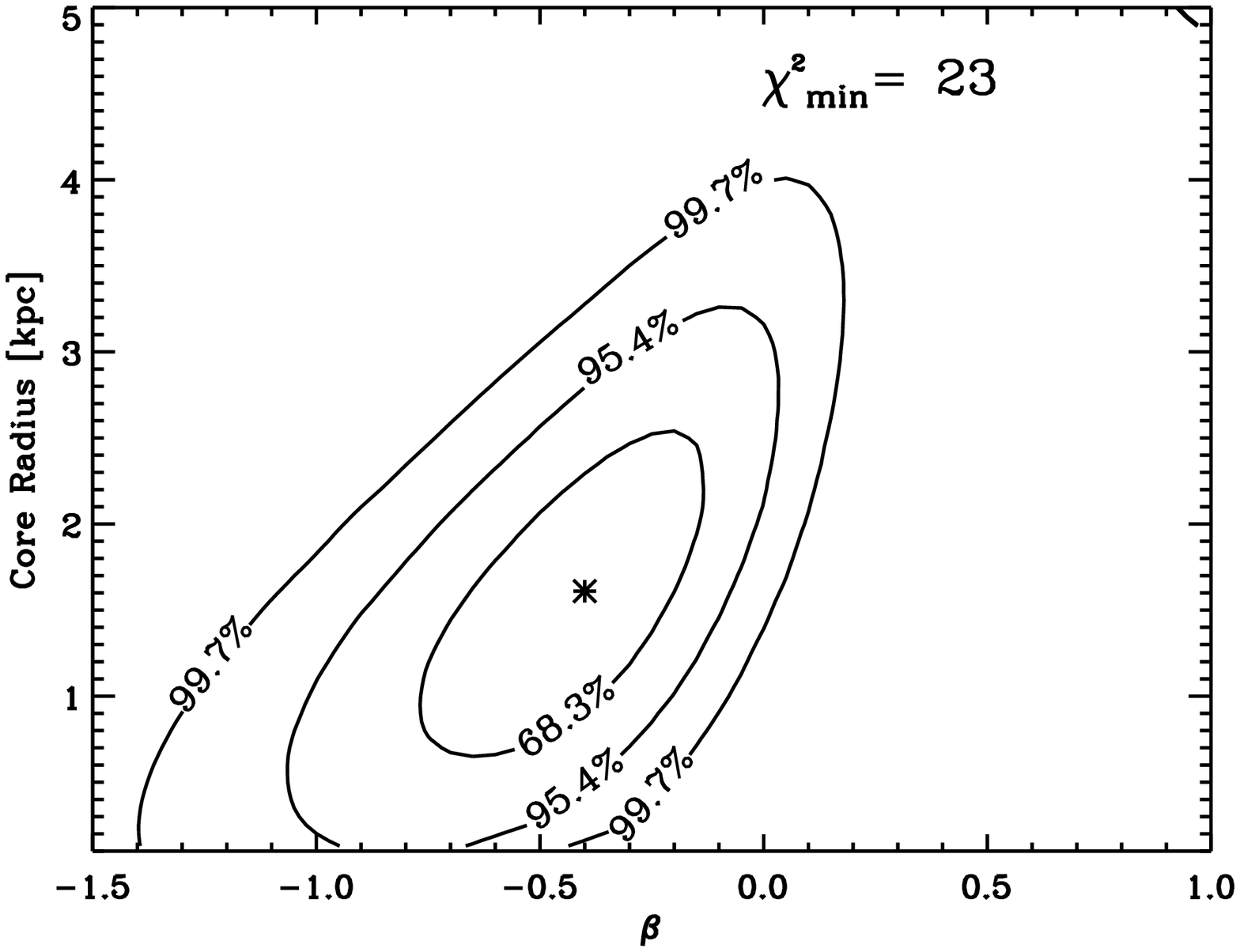}
\includegraphics[width=85mm]{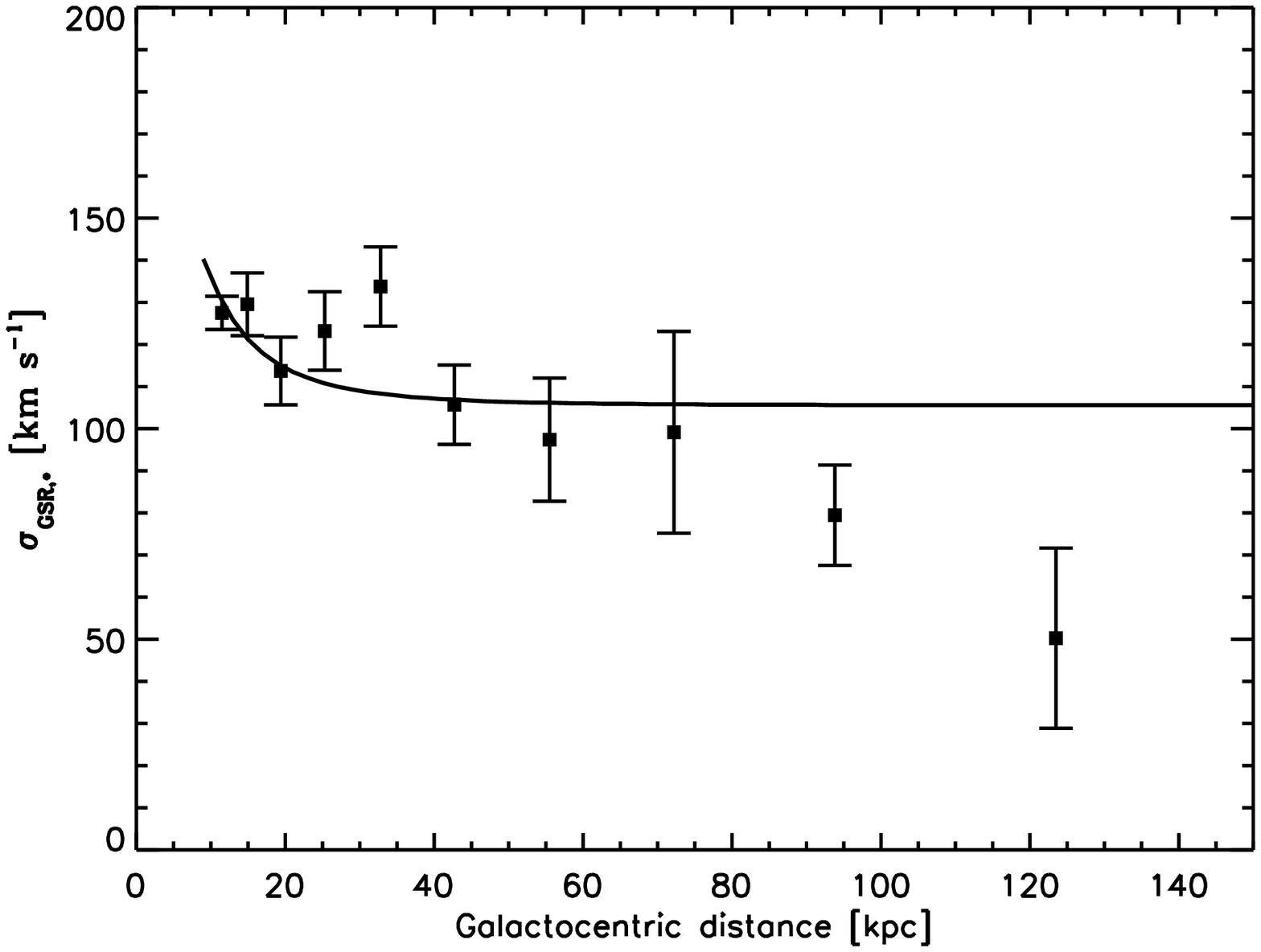}
\caption{Left: Contour plot of $\Delta{\chi}^2$ corresponding to a
probability of the 68.3\%, 95.4\%, 99.7\% (1$\sigma$, 2$\sigma$,
3$\sigma$) for the isothermal sphere model with constant
anisotropy. The asterisk indicates the location of the minimum
$\chi^2$ (whose value is shown in the upper right corner).  Right:
Observed radial velocity dispersion (squares with error-bars) overlaid
on the best fit model for the isothermal mass distribution (solid
line).  }
\label{fig:bestfit_iso}
\end{center}
\end{figure*}  

\item {\it Truncated Flat model.} This density profile was recently
introduced by W\&E99
to describe the dark matter
halo of Local Group galaxies. It is a mathematically convenient
extension of the Jaffe (1983) model. The form of the density profile
of the Truncated Flat model (hereafter TF) is
\begin{equation}
\rho(r) = {M\over{4\pi}} {a^2\over{r^2(r^2+a^2)}^{3/2}}
\end{equation}
where $a$ is the scale length and $M$ the total mass of the system.
For $r \gg a$, the density falls off as $\rho\propto r^{-5}$. The
circular velocity due to this density distribution is
\begin{equation}
V_{\rm c}^2(r) =\frac{V_0^2 a}{(r^2 + a^2)^{1/2}}.
\end{equation}
We set $V_0 = $ 220 \kms (W\&E99). The resulting rotation curve is
flat in the inner part, with amplitude $V_0 = \sqrt{GM/a}$, and
becomes Keplerian for $r \gg a$. Having fixed the amplitude of the
circular velocity ($V_0$), this model is reduced to a one parameter-family
characterized by the scale length $a$, or the mass $M$.
\end{itemize}

\subsection{Results}
\label{sec:results}

\subsubsection{Models with constant velocity anisotropy}

\begin{figure*}
\begin{center}
\includegraphics[width=80mm]{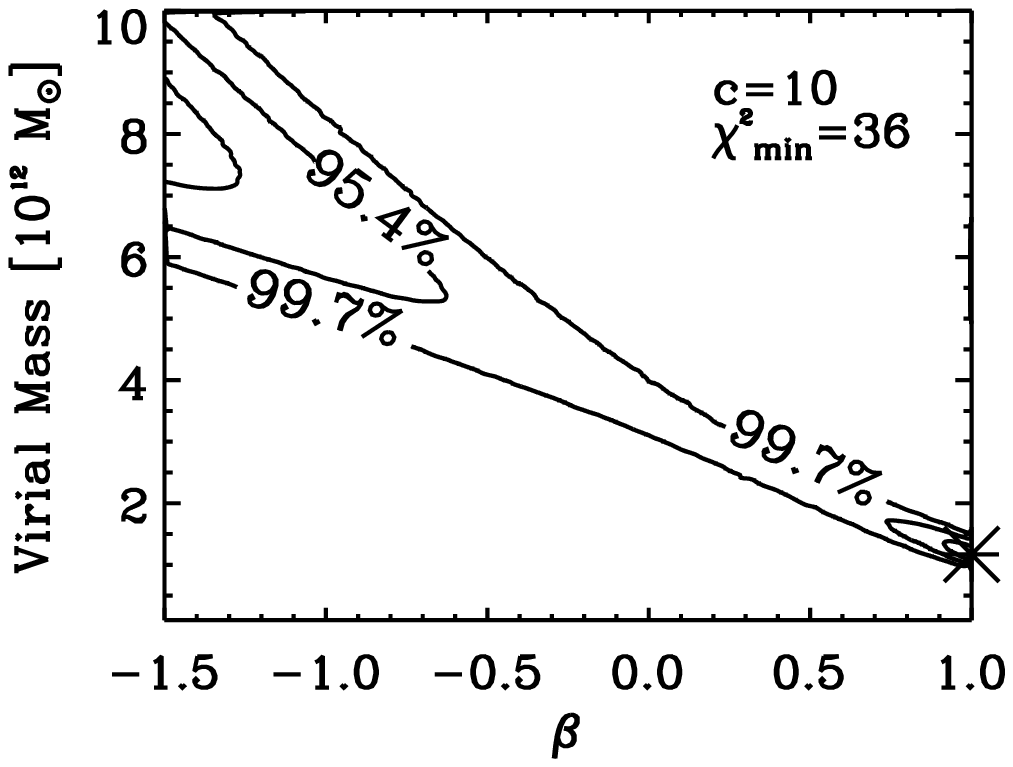}
\includegraphics[width=80mm]{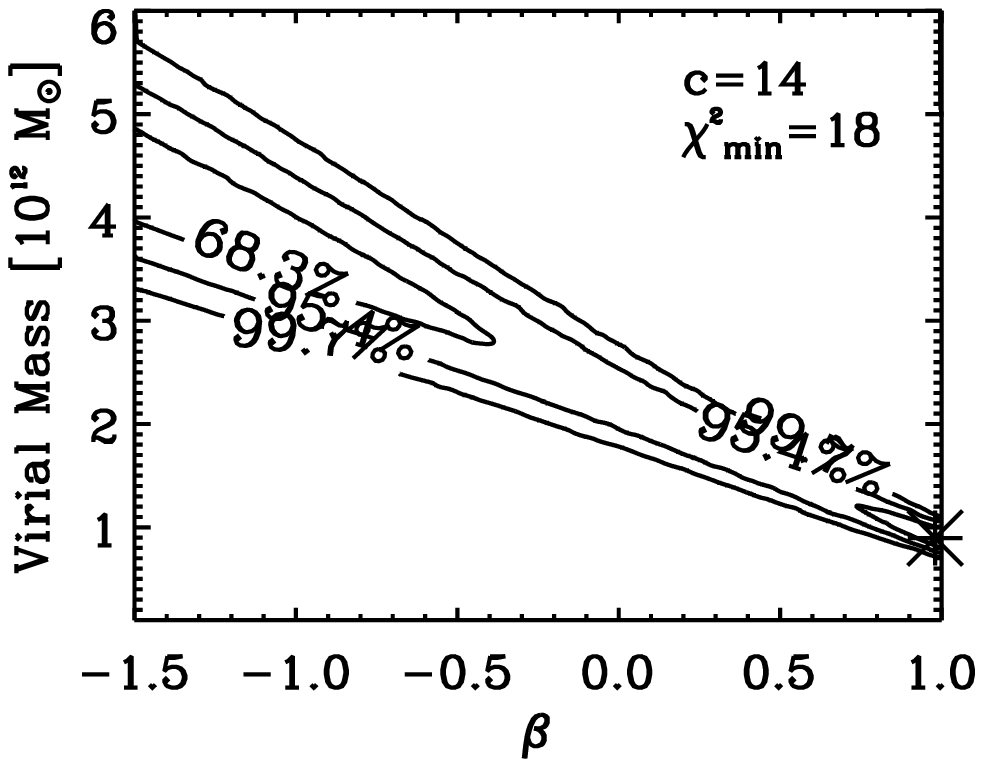}
\includegraphics[width=80mm]{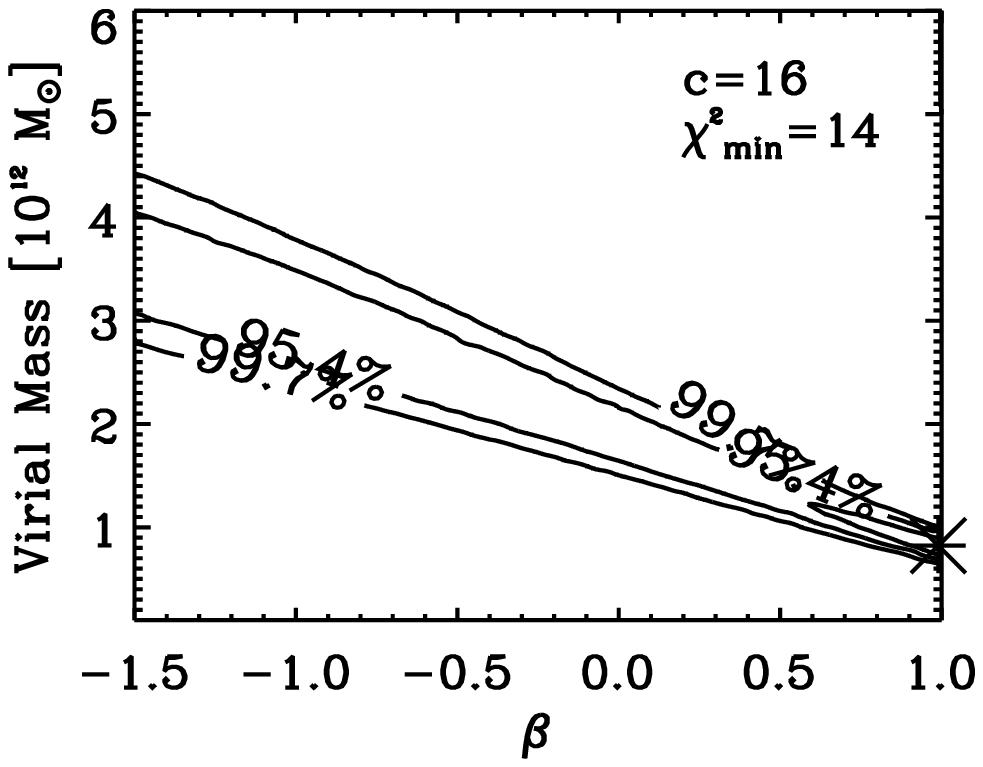}
\includegraphics[width=80mm]{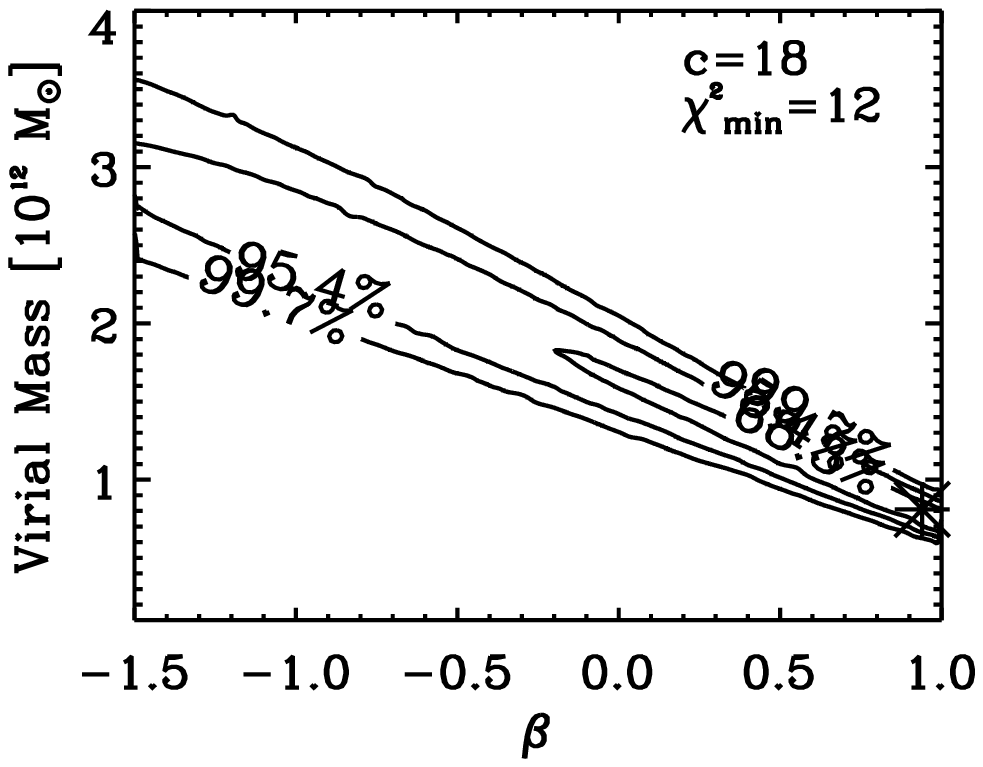}
\includegraphics[width=140mm]{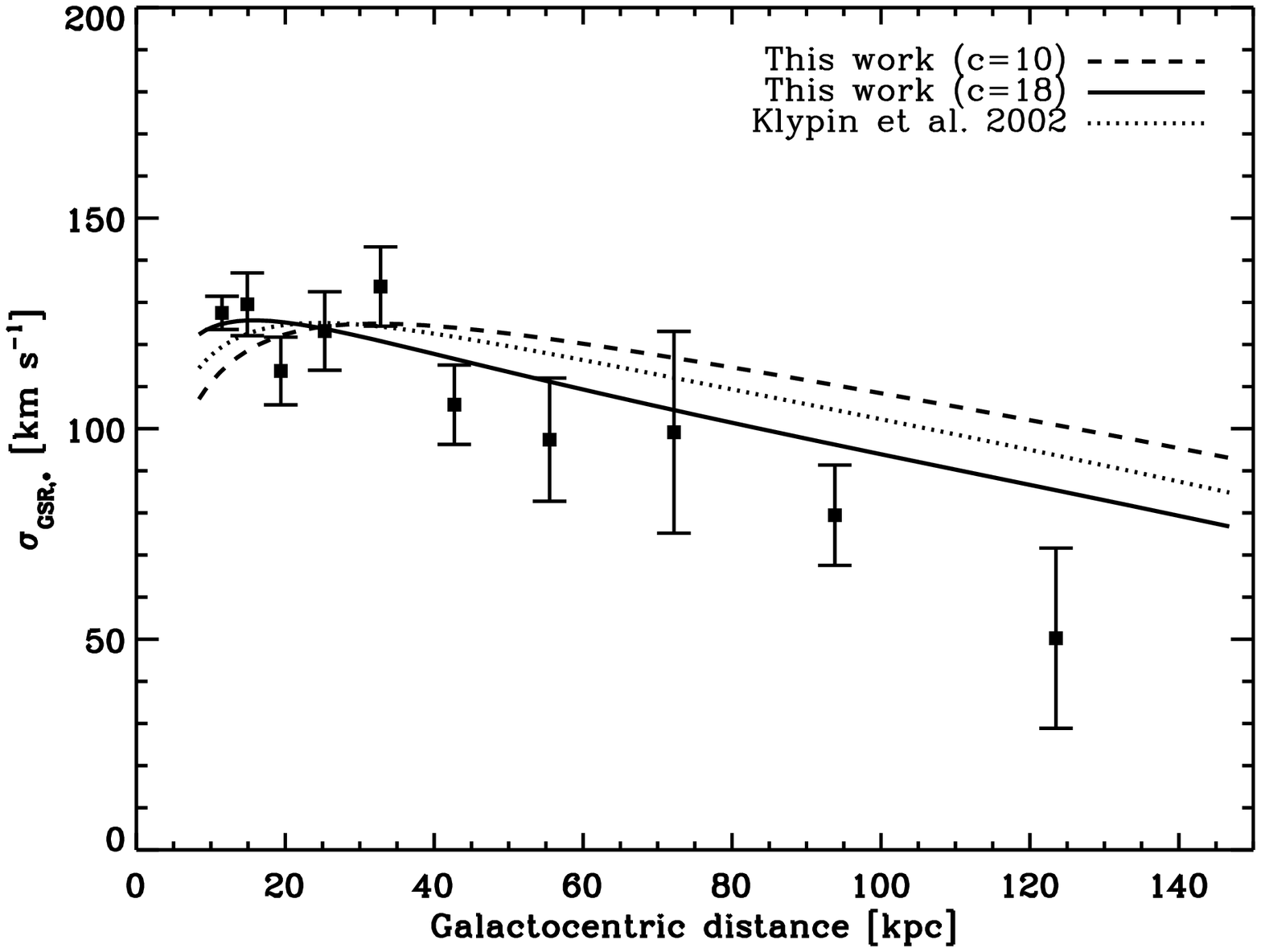}
\caption{ Top: Contour plot of $\Delta{\chi}^2$ corresponding to a
probability of the 68.3\%, 95.4\%, 99.7\% (1$\sigma$, 2$\sigma$,
3$\sigma$) for the NFW model at four different concentrations. The
value of the concentration and minimum ${\chi}^2$ are shown in the
upper right corner of each panel. The asterisk indicates the location
of the minimum $\chi^2$ and hence of the best-fitting parameters. 
The virial mass is given in units of 10$^{12}$ \sm. Bottom:
Observed radial velocity dispersion (squares with error-bars) overlaid
on two of the best fit models for the NFW mass distributions (dashed 
line: c=10; solid line: c=18).  The dotted curve corresponds to the
Galactocentric radial velocity dispersion profile obtained using the
preferred model (B1) of Klypin et al. (2002).}
\label{fig:bestfit_nfw}
\end{center}
\end{figure*}

\begin{figure*}
\begin{center}
\includegraphics[width=85mm]{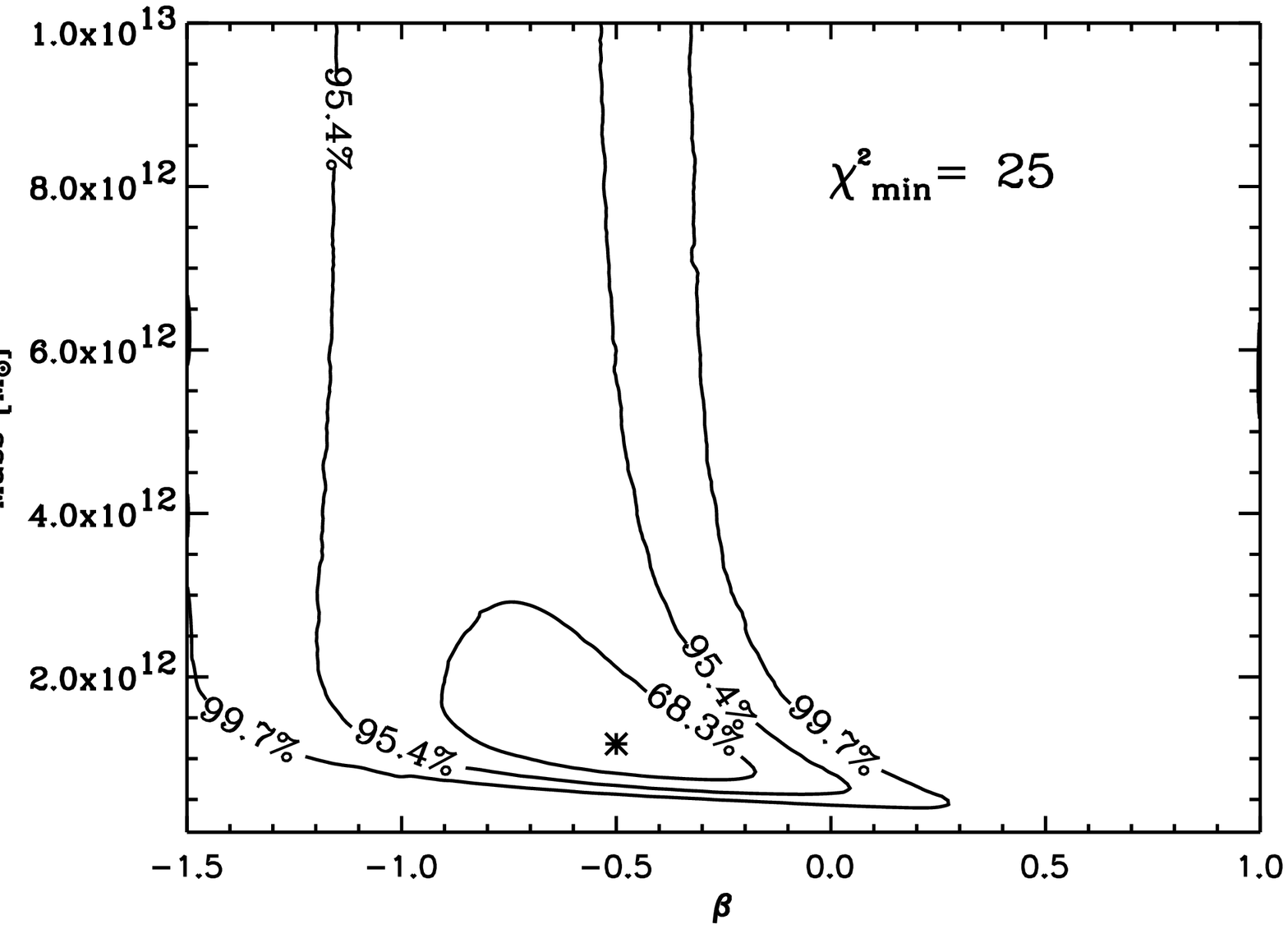}
\includegraphics[width=85mm]{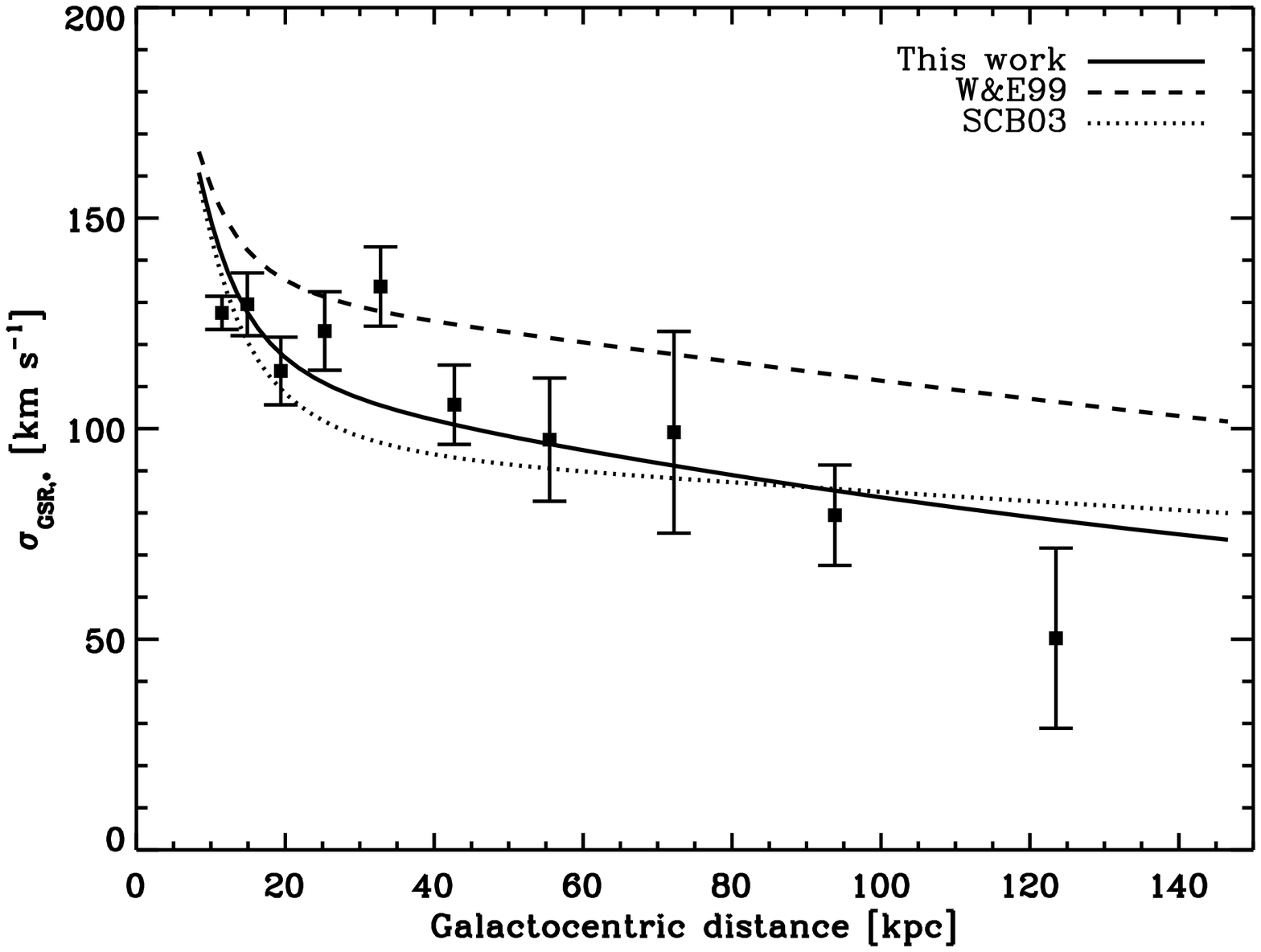}
\caption{Left: Contour plot of $\Delta{\chi}^2$ corresponding to a
probability of the 68.3\%, 95.4\%, 99.7\% (1$\sigma$, 2$\sigma$,
3$\sigma$) for the TF model. The asterisk indicates the location of
the minimum $\chi^2$ (whose value is shown in the upper right corner).
Right: Observed radial velocity dispersion (squares with error-bars)
overlaid on the best fit model for the TF mass distribution (solid
line). The dashed line shows the Galactocentric radial velocity
dispersion obtained using the best-fitting parameters from previous works
(dashed: W\&E99; dotted: SCB03).  }
\label{fig:bestfit_tf}
\end{center}
\end{figure*}  

The methodology we use consists in comparing the measured Galactocentric 
radial velocity dispersion $\sigma_{\rm GSR,*}$ for each of the distance bins
with that predicted for the different models discussed in
Sec.~\ref{sec:models}. For the latter, we explore the space of
parameters which define each model and determine the $\chi^2$ as:
\begin{equation}
\chi^2 = \sum_{i=1}^{Nbins} \biggl(\frac{\sigma_{{\rm GSR}_i,*} - 
\sigma_{ GSR,*}(r_i;
\beta, p)}{\epsilon_r}\biggr)^2.
\end{equation}
Here, the variable $p$ denotes a characteristic parameter of each
model (e.g. scale length or total mass), while $\epsilon_r$ is the
error in the observed radial velocity dispersion as estimated through
the bootstrap sampling technique described before. The best-fitting 
parameters are defined as those for which $\chi^2$ is minimized. 

In the case of the isothermal sphere, the free parameters are the dark
matter halo core radius, $r_{\rm c}$, and the stellar velocity
dispersion anisotropy parameter, $\beta$. The left panel of
Fig.~\ref{fig:bestfit_iso} shows the $\chi^2$ contours for this model.
The minimum $\chi^2$ value is $\chi_{\rm min}^2=23$ for $r_{\rm c}$ =
1.6 kpc and $\beta = -0.4$, with 1-$\sigma$ contours encompassing
$0.6\lesssim r_{\rm c}\lesssim2.6$ and
$-0.7\lesssim\beta\lesssim-0.1$. This corresponds to a best-fitting
mass $M= 1.3 \times10^{12}$ \sm (note that, since the mass for the
pseudo-isothermal model is not finite, we quote the mass within our
last measured point, at $r=$ 120 kpc). The 1-$\sigma$ errors on the
mass, calculated from the 1-$\sigma$ errors for the core radius, lead
to a relative error of the order of 1\%. The reason for this small
value is due to the fact that the best-fitting core radius is very
small, and hence variations in its value (even by 100\%) will barely
affect the mass enclosed at large radii.  On the right panel of
Fig.~\ref{fig:bestfit_iso} we plot the Galactocentric radial velocity
dispersion for this best-fitting model. As expected, this model
predicts a velocity dispersion that is roughly constant with
radius. However, the observed $\sigma_{\rm GSR,*}$ shows a rather
strong decline at large radii, which is not reproduced by the
pseudo-isothermal halo model.

The top panels of Fig.~\ref{fig:bestfit_nfw} show the $\chi^2$
contours for the NFW model for 4 different concentrations ($c=10$, 14,
16, and 18). Note that the minimum $\chi^2$ value decreases for
increasing concentrations. Since the concentration is defined as
$c=r_{\rm v}/r_{\rm s}$, for a fixed mass (or virial radius $r_{\rm
v}$) a larger $c$ implies a smaller scale radius. This results in a
radial velocity dispersion that starts to decline closer to the centre
in comparison to a halo of lower concentration, reproducing better the
trend observed in the data. Our $\chi^2$ fitting technique yields for
$c=10$ a best-fitting virial mass of 1.2$\times10^{12}$ \sm
($\chi_{\rm min}^2= 36$), while for $c=18$, $M_{\rm v} =
0.8\times10^{12}$ \sm ($\chi_{\rm min}^2=12$).  We find that the
velocity anisotropy for the minimum $\chi^2$ is almost purely radial
in all cases. In the bottom panel of Fig.~\ref{fig:bestfit_nfw} we
show the observed Galactocentric radial velocity dispersion overlaid
on two of the best-fitting NFW models. Note that beyond 40 kpc, the
model with $c=10$ is clearly inconsistent with the data at the
1$\sigma$ level at $r\sim$ 40 and 50 kpc and at the 2$\sigma$ level in
the last two bins. On the other hand, the $c=18$ model gives a good
fit of the data out to 30 kpc but overpredicts the velocity dispersion
at large radii at the 1$\sigma$ level.  We thus consider the NFW model
with $M_{\rm v}= 0.8^{+1.2}_{-0.2}\times 10^{12}$\sm and $c=18$ as
producing the best fit.  Fig.~\ref{fig:bestfit_nfw} also shows the
favourite model of Klypin et al. (2002) with $M_{\rm
v}=1.0\times10^{12}$ and $c=$12 (dotted curve). Since no velocity
anisotropy was given in the source we performed a $\chi^2$ fit to our
data using the parameters from Klypin et al. (2002) and leaving
$\beta$ as a free parameter. This favoured once again an almost purely
radial anisotropy. The fit obtained in this case is very similar to
that found in our $c=10$ model.

Since our last measured point is at $r_{\rm last}\sim$ 120 kpc, the
constraining power of our data is stronger in the region enclosed by
this radius. The value of the virial mass we just derived is an
extrapolation of the model at larger distances. For completeness, we
quote here the mass within 120 kpc for our best fitting NFW model with
$c=18$, $M(<120\,{\rm kpc})= 5.4^{+2.0}_{-1.4}\times 10^{11}$\sm (the
errors are calculated from the 1$\sigma$ errors in the best-fitting
mass).

The left panel of Fig.~\ref{fig:bestfit_tf} shows the contour plot for
the TF model. Our best fit has a mass of 1.2$^{+1.8}_{-0.5}\times
10^{12}$\sm and $\beta=-0.50\pm0.4$ (${\chi}^2_{\rm min}=$ 25). The
mass enclosed in 120 kpc is $M(<120\, {\rm kpc})=
9.0^{+6.0}_{-3.0}\times 10^{11}$\sm. 
Our results are compatible with the work of W\&E99:
they find a mass of $M=1.9^{+3.6}_{-1.7}\times10^{12}$\sm, even
though they favour a slightly radially anisotropic velocity ellipsoid.
The right panel of Fig.~\ref{fig:bestfit_tf} shows the data overlaid
on our best-fitting model (solid line).  Visual inspection shows that the
large value obtained for the minimum ${\chi}^2$ is driven by the
discrepancy between model and data in the bins at 11.5 and 33 kpc.
However, at large radii our TF model with $M= 1.2\times 10^{12}$\sm
provides a good representation of the data.  Figure
\ref{fig:bestfit_tf} also shows that the favourite W\&E99 model
(dashed curve), having a larger mass and a more radially anisotropic
velocity ellipsoid, overpredicts the Galactocentric radial velocity
dispersion. On the other hand, the TF model of SCB03, for which
$M=2.5\times10^{12}$\sm and $\beta=-1.25$, i.e. heavier halo whose
ellipsoid is much more tangentially anisotropic, declines too quickly
in the inner part and tends to flatten at large radii (dotted curve),
not following the trend shown by the data.

The comparison of the fits produced by the constant anisotropy TF and
NFW models shows that the latter reproduces better the trends in the
data as a whole, from small to large radii. However, at very large
radii it tends to overpredict the velocity dispersion. In this regime,
the TF model provides a much better fit. This can be understood as
follows.  In the region between 50 and 150 kpc, where $\sigma_{\rm
GSR,*}$ shows the decline, the slope of the TF model ranges between
$-3$ and $-4$ whilst the slope of the NFW density profile is around
$-2.5$.  This means that, in models with a constant velocity
anisotropy, a steep dark matter density profile at large radii is
favoured by the data.  

\subsubsection{Toy models for the velocity anisotropy}

We will now briefly relax the assumption that $\beta$ is constant with
radius.  We shall explore the following models for $\beta(r)$:
\begin{itemize}
\item {\it Model $\beta$-rad (Radially anisotropic).} Diemand, Moore \& Stadel
(2004) have found in N-body $\Lambda$CDM simulations that the
anisotropy of subhalos velocities behaves as
\begin{equation}
\beta(r) \simeq 0.35\frac{r}{r_{\rm v}}, \qquad \textrm{for}\,\, r \le
r_{\rm v}.
\label{eq:beta_r}
\end{equation}
We will use this cosmologically motivated functional form to study the
effect of an increasingly radially anisotropic velocity ellipsoid in
our modelling of the radial velocity dispersion curve.

\item {\it Model $\beta$-tg (Tangentially anisotropic).} Proper motion
measurements of the Magellanic Clouds and Sculptor, Ursa Minor, and
Fornax dwarf spheroidals suggest that the tangential velocities of
these objects are larger than their radial motions (Kroupa \& Bastian
1997; Schweitzer et al. 1995, 1997; Dinescu et al. 2004). If
confirmed, this would have as consequence that the velocity ellipsoid
should be tangentially anisotropic at large radii. To explore the
effect on our dynamical models of a velocity ellipsoid that becomes
increasingly more tangential, we consider the following toy-model:
\begin{equation}
\beta(r)= \beta_0  - \frac{r^2}{h^2},
\label{eq:beta_t}
\end{equation}
where we set the scale factor $h=$ 120 kpc. We choose two values for
$\beta_0$: in the first case we arbitrarily fix it to 1 (model
$\beta$-tg$_{\rm toy}$); in the second model ($\beta$-tg$_{\rm SN}$)
we use a a sample of 91 nearby halo stars from Beers et al. (2000) --
within 0.5 kpc from the Sun and with [Fe/H]$<-$1.5 -- to normalize our
model.  In this case, we find that $\beta(R_{\odot})= 0.33$ and
therefore, $\beta_0= 0.33 + R_{\odot}^2/h^2$.

\begin{figure}
\begin{center}
\includegraphics[width=80mm]{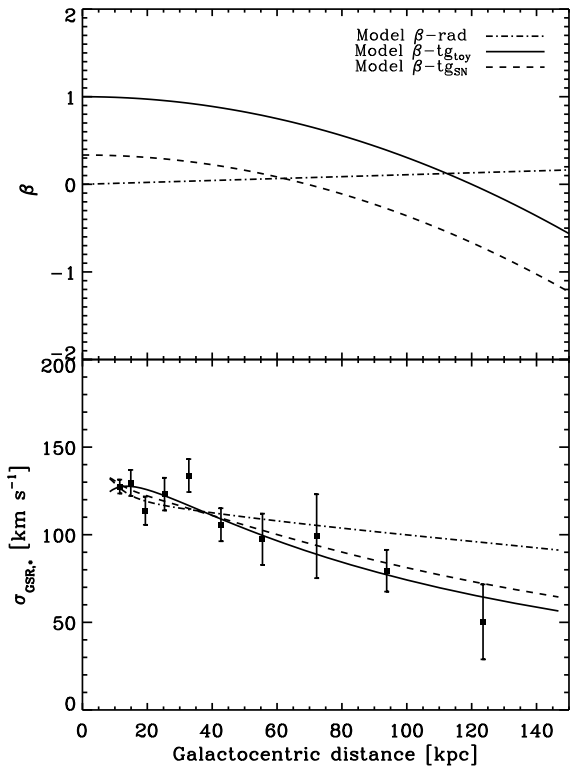}
\caption{Top: The solid and dashed curves correspond to two toy-models
for a velocity ellipsoid that becomes more tangentially anisotropic
with radius. The dashed-dotted line shows a model for an increasingly
radially anisotropic ellipsoid from Diemand, Moore \& Stadel (2004).
Bottom: best-fitting models for an NFW halo of concentration $c=18$
corresponding to the $\beta$ profiles shown in the top panel. }
\label{fig:beta}
\end{center}
\end{figure}

\end{itemize}   

Using the models for $\beta(r)$ described above, we perform again the
${\chi}^2$ best-fitting procedure for an NFW model of $c=18$. There
is, therefore, in all cases, only one free parameter: the virial mass.
The results of this new analysis are shown in the bottom panel of
Fig.~\ref{fig:beta}. The $\beta$-$rad$ model, for which the velocity
ellipsoid becomes more radially anisotropic with radius, has
$\chi_{\rm min}^2=15$. Even though the predicted radial velocity
dispersion of this model does decrease with radius, this decline is of
insufficient amplitude to reproduce the trend shown by the data. Note
that this model, motivated by dark-matter simulations, provides a
poorer fit than the constant $\beta$ model. Models where the velocity
ellipsoid becomes more tangentially anisotropic with radius,
$\beta$-tg$_{\rm toy}$ and $\beta$-tg$_{\rm SN}$, follow very well the
data, and have $\chi_{\rm min}^2= 6$ and 7, respectively. We find
that, for model $\beta$-tg$_{\rm toy}$, the best-fitting virial mass
is $M_{\rm v}=$ 8.8 ($\pm$ 0.7, $\pm$ 1.2)$\times10^{11}$\sm (at the
1$\sigma$, 2$\sigma$ level), and $M_{\rm v}=$1.5 ($\pm$ 0.1, $\pm$
0.2)$\times10^{12}$\sm (at the 1$\sigma$, 2$\sigma$ level) for
$\beta$-tg$_{\rm SN}$. For the $\beta$-tg$_{\rm toy}$ model, we find
that mass enclosed in 120 kpc is $M(<120\, {\rm kpc}) = 5.9\pm 0.5
\times 10^{11}$\sm; for the $\beta$-tg$_{\rm SN}$ model, $M(<120\,
{\rm kpc}) = 9.0\pm 0.6 \times 10^{11}$\sm. Table~2 summarizes the
best-fitting parameters for our favourite models.

This analysis highlights the mass-anisotropy degeneracy, since it
shows that, even for the same functional form of $\beta$, the
best-fitting value of the virial mass can differ by a factor of
two. Note that the best-fitting values of the virial mass for the
$\beta$-tg$_{\rm toy}$ model and the $\beta = cst$ are very
comparable, but this is a reflection of the fact that the two
anisotropy parameters are not too similar throughout a fair range of
the distances probed by the sample. However, since the value of
$\beta$ in the Solar neighbourhood is in the range 0.5$\pm$0.1 
(Chiba \& Yoshii 1998), this
would tend to suggest that, given that the ellipsoid needs to be
tangentially anisotropic at large radii to give a good fit to the
data, a higher value of the total mass is more likely.

If we apply the same kind of analysis to the pseudo-isothermal sphere
mass model, it is clear that $\beta$ has to decrease more strongly
with radius than the above $\beta$-tg model used in combination with
the NFW profile in order to give a reasonable fit to the data. This is
in line with the results of Sommer-Larsen et al. (1997). By assuming a
logarithmic potential for the dark matter halo, they found a velocity
ellipsoid radially anisotropic at the Solar circle ($\beta\sim$ 0.5)
and tangentially anisotropic for $r\gtrsim$ 20 kpc. At $r \sim$ 50 kpc
the expected value of $\beta\sim- 1$.  The Sommer-Larsen et al. (1997)
model is consistent with our findings out to $\sim$ 50 kpc; however,
if we extrapolate the predicted trend for $\beta$ to larger
Galactocentric distances, we notice that $\beta$ does not decrease
sufficiently rapidly to explain the decline observed in our data (see
also Appendix B).

From the above analysis it is evident that assumptions on $\beta$ for
a particular mass model, can strongly influence the performance of the
mass model. However, not all functional forms of $\beta$ for a given
mass model produce a good fit to the data. More accurate proper motion
measurements for a larger number of halo tracers and covering a larger
range in Galactocentric distances will enable us to understand which
trend in radius $\beta$ is following and, therefore, to establish more
uniquely which mass model is preferred by the data.

In addition to varying the velocity anisotropy parameter $\beta$ as
function of radius, it is also possible to consider the effect of
changing the slope $\gamma$ of the stellar density profile of the
Galactic halo. In this case, however, the data is much more
restrictive in the choice of possible models, since it is well-known
that $\gamma \sim 3 - 3.5$ out to $\sim 50$ kpc (Yanny et al. 2000).
Equation (2) shows that possible variations of the stellar halo power
law $\gamma$ with radius can ``conspire'' with variations of $\beta$
to reproduce the same radial velocity dispersion profile. We examine
this issue further in Appendix~B.

\begin{table*} 
\begin{center}
\begin{tabular}{lccccccc}
\hline & $\chi_{\rm min}^2$ & $\beta$ & Mass [10$^{12}$\sm] & scale
length [kpc] \\ 
\hline TF model & 25 & $-$0.5 & $1.2_{-0.5}^{+1.8}$ & 105\\ 
\hline NFW model & 12 & 0.94 & $0.8_{-0.2}^{+1.2}$ & 255 (c = 18) \\ 
\hline NFW model ($\beta$-tg$_{\rm SN}$) & 7 & $\beta'(r) < 0$ & 1.5$\pm$0.1 & 312 (c = 18)\\ 
\hline
\end{tabular}
\caption{Values of the parameters for our favourite best-fitting models; the
scale length corresponds to 
{\it a} for the TF model and $r_{\rm v}$ for the NFW. 
} \label{tab:tab2}
\end{center}
\end{table*}

\section{Discussion and conclusions}
\label{sec:disc}

We have derived the radial velocity dispersion profile of the stellar
halo of the Milky Way using a sample of 240 halo objects with accurate
distance and radial velocity measurements. The new data from the
``Spaghetti'' Survey led to a significant increase in the number of
known objects for Galactocentric radii beyond 50 kpc, which allowed a
more reliable determination of the dispersion profile out to very
large distances. Our most distant probes are located at $\sim 120$
kpc, which in comparison to previous works (e.g. Sommer-Larsen et
al. 1997) corresponds to an increase of 70 kpc in probing the outer
halo. The Galactocentric radial velocity dispersion measured is
approximately constant ($\sigma_{\rm GSR,*} \sim$ 120 \kms) out to 30
kpc (consistent with Ratnatunga \& Freeman 1986) and then it shows a
continuous decline out to the last measured point (50 $\pm$ 22 \kms at
120 kpc). This fall-off has important implications for the density
profile of the dark matter halo of the Milky Way. In particular, in
the hypothesis of a constant velocity anisotropy, an isothermal sphere
can be immediately ruled out as model for the Galactic dark halo as
this predicts a nearly constant radial velocity dispersion curve.

We have also considered two other possible models for the dark halo: a
truncated flat (TF) and a Navarro, Frenk \& White (NFW) profile. We
have compared the radial velocity dispersion observed with that
predicted in these models for a tracer population (stellar halo) 
embedded in a potential provided by the dark halo. By
means of a $\chi^2$ test, we were able to derive the characteristic
parameters and velocity anisotropy of these models that are most
consistent with the observed data.  

In the case of a TF profile, the favourite model for the Milky Way
dark matter halo has a mass $ M = 1.2^{+1.8}_{-0.5}\times10^{12}$\sm,
with a corresponding velocity anisotropy $\beta= -0.50\pm0.4$.  The
data are also compatible with an NFW dark halo of $M_{\rm v} =
0.8^{+1.2}_{-0.2}\times10^{12}$ \sm and $-$0.3$\lesssim \beta \lesssim
1$ for a concentration $c=$18. The comparison of the fits produced by
the constant anisotropy TF and NFW models shows that the latter
reproduces better the trends in the data as a whole, from small to
large radii. However, at very large radii it tends to overpredict the
velocity dispersion. In this regime the TF model --having a steeper
density profile-- provides a much better fit.

Our determination of the dark halo mass of the Milky Way is consistent
with previous works: the preferred TF model of W\&E99 gives a mass $M
= 1.9\times 10^{12}$\sm, with a 1-$\sigma$ range of $0.2 <
M[10^{12}$\sm$] < 5.5$ and $-0.4 < \beta < 0.7$; the favourite model
from Klypin et al. 2002 gives $M=1.0\times 10^{12}$ \sm with $c=$12.
However, the radial velocity dispersion predicted by these two models
is larger than the observed one.  The discrepancy between the observed
low values of the radial velocity dispersion at large radii and that
predicted for heavy dark halos raises the question of whether the
velocity dispersion in the two most distant bins may be affected by
systematics, such as the presence of streams, which could lower their
values.

The two bins in question are centered at $\sim 90$ kpc and $\sim 120$
kpc, and contain 6 and 3 objects respectively: 4 satellite galaxies
and 5 globular clusters. The minimum angular separation of any two
objects in these bins is 40$^{\circ}$, for the satellites, and
49$^{\circ}$, for the GCs.  When considering the sample with 9 objects
only two of these objects appear to be close on the sky: one globular
cluster and one satellite galaxy that are located at $(l,b) \sim$
(241$^{\circ}$,42$^{\circ}$). Although these are at similar distances
of 96 kpc and 89 kpc, respectively, their line of sight radial
velocities differ by more than 140 \kms, thus making any physical
association extremely unlikely.

We have also investigated the effect of a velocity anisotropy that
varies with radius on the velocity dispersion $\sigma_{\rm GSR,*}$ in
the case of an NFW halo of concentration $c=$18. We find that the
velocity anisotropy, which is radial at the Solar neighborhood, needs
to become more tangentially anisotropic with radius in order to fit
the observed rapid decline in $\sigma_{\rm GSR,*}$.  In the case of an
isothermal dark matter halo, the $\beta$ profile needs to decline even
more steeply than in the NFW case in order to fit the data.

We conclude that the behaviour of the 
observed velocity dispersion can be explained either by  
a dark matter halo following a steep density profile at large radii and 
constant velocity anisotropy, or by a halo with a less steep profile 
whose velocity ellipsoid becomes tangentially anisotropic at large radii. 
In order to distinguish between an NFW profile 
and a TF model, proper motions are fundamental since they enable the 
direct determination of the velocity anisotropy profile. Proper motions of 
GCs and satellites are becoming available (Dinescu et al. 1999; 
Piatek et al. 2003; Dinescu et al. 2004) 
albeit with large errors because of the very distant location of these objects.
We may have to wait until Gaia is launched to determine the density profile 
of the Galactic dark-matter halo. 

\section*{Acknowledgments}

Giuseppina Battaglia gratefully acknowledges Eline Tolstoy.  We thank
Simon White and the anonymous referee for useful suggestions.  This
research has been partially supported by the Netherlands Organization
for Scientific Research (NWO), and the Netherlands Research School for
Astronomy (NOVA). Heather Morrison, Edward W.~Olszewski and Mario
Mateo acknowledge support from the NSF on grants AST 96-19490, AST
00-98435, AST 96-19524, AST 00-98518, AST 95-28367, AST 96-19632 and
AST 00-98661.
We used the web catalog http://physwww.mcmaster.ca/$\sim$harris/mwgc.dat 
for the Milky Way globular cluster data.

\section*{Appendix A}

The Galactocentric radial velocity $v_{\rm GSR}$ (i.e. the l.o.s. 
heliocentric velocity $V_{\rm los}$ corrected for the solar motion and 
LSR motion)
is related to the true Galactocentric radial, $v_r$, and tangential, 
$v\rm{_t}$, velocity by
\begin{equation}
v_{\rm GSR} = v_r \, \hat{\epsilon_r} \cdot \hat{\epsilon\rm{_R}} +
v\rm{_t} \, \hat{\epsilon\rm{_t}} \cdot \hat{\epsilon\rm{_R}} 
\end{equation}
where $\hat{\epsilon_r}$ is the unit vector in the radial
direction towards the object as seen from the Galactic centre,
$\hat{\epsilon\rm{_t}}$ is the unit vector in tangential direction in
the same reference frame, and $\hat{\epsilon\rm{_R}}$ is the unit
vector in the radial direction from the Sun to the object. The two
scalar products depend on the heliocentric and galactocentric
distances ($d$ and $r$) and position on the sky of the object ($\phi$,
$\theta$).  For a given distribution function $f(\bar{r},\bar{v})$,
the velocity dispersion profile (seen from the Sun) is given by
$\sqrt{\langle v_{\rm GSR}^2\rangle}$, and can be found by squaring
Eq.~(15) and integrating over all the velocities and averaging over
the solid angle:
\begin{eqnarray*}
\langle v_{\rm GSR}^2\rangle|_{\rm{\Omega -av}} =    
 \frac{1}{\D \int d^2 \Omega} 
\Big[ \int d^2\Omega \, k(r,\theta,\phi)\, 
\int d^3v \, v_r^2 \, f(\bar{r},\bar{v})  + \\
  \int d^2\Omega \, h(r,\theta,\phi)\, 
\int d^3v \, v_{\rm t}^2 \, f(\bar{r},\bar{v})  \Big], 
\end{eqnarray*}
or
\begin{eqnarray}
\langle {v_{\rm GSR}^2\rangle|_{\rm{\Omega -av}} =   
\frac{1}{4\pi}\Big(\int d^2\Omega \, k(r,\theta,\phi)\, 
 \langle v_r^2\rangle +{}} 
                        \nonumber\\
{}+\int d^2\Omega \, h(r,\theta,\phi)\,
\langle v_{\rm t}^2\rangle \Big),
\label{eqn:omega_av}
\end{eqnarray}
where we have defined
\[
\hat{\epsilon\rm{_R}}= \frac{\bar{r}-\bar{R}_{\odot}}{d},
\]
\[
k(r,\theta,\phi)=(\hat{\epsilon_r} \cdot \hat{\epsilon\rm{_R}})^2= 
\Big(\frac{r+R_{\odot}\,\cos\phi\, \sin\theta}{d}\Big)^2,
\]
and 
\[
h(r,\theta,\phi)=(\hat{\epsilon\rm{_t}} \cdot \hat{\epsilon\rm{_R}})^2= 
\frac{{R_{\odot}}^2}{d^2} \, (\cos^2\theta \cos^2\phi + \sin^2\phi).
\]
Eq.~(\ref{eqn:omega_av}) can thus be expressed as
\[
\langle v_{\rm GSR}^2\rangle|_{\rm{\Omega -av}} = \langle
v_r^2\rangle K(r) + \langle v_{\rm t}^2\rangle H(r).
\]
If we assume that $\langle v_\theta^2\rangle= \langle
v_\phi^2\rangle$, and from the definition of the velocity anisotropy
$\beta$ we find $\langle v_{\rm t}^2\rangle= 2 \langle v_r^2\rangle
(1-\beta )$, then it follows that
\begin{equation}
\langle v_{\rm GSR}^2\rangle|_{\rm{\Omega -av}}= 
\langle {v_r}^2\rangle \, [K(r) + 2 (1-\beta )H(r)].
\end{equation}
By assuming $\langle v_{r}\rangle=0$ 
and $\langle v{\rm_{t}}\rangle=0$, it follows that 
$\langle v{\rm_{GSR}}\rangle=0$;
by performing the above integrals for $r > R_\odot$, we find that the
Galactocentric radial velocity dispersion is related to the true
radial velocity dispersion by
\begin{equation}
{\sigma{\rm_{GSR}}}(r) = \sigma_{r}(r)\,\sqrt{1 + 2\,
( 1 - \beta) H(r)},
\end{equation}
where
\begin{equation}
H(r) = \frac{r^2 + R_{\odot}^2}{4r^2} - \frac{{(r^2-
R_{\odot}^2)}^2}{8r^3 R_{\odot}}{\rm ln}\frac{r+R_{\odot}}{r-R_{\odot}}.
\end{equation}

\begin{figure*}
\includegraphics[width=0.24\textwidth,clip]{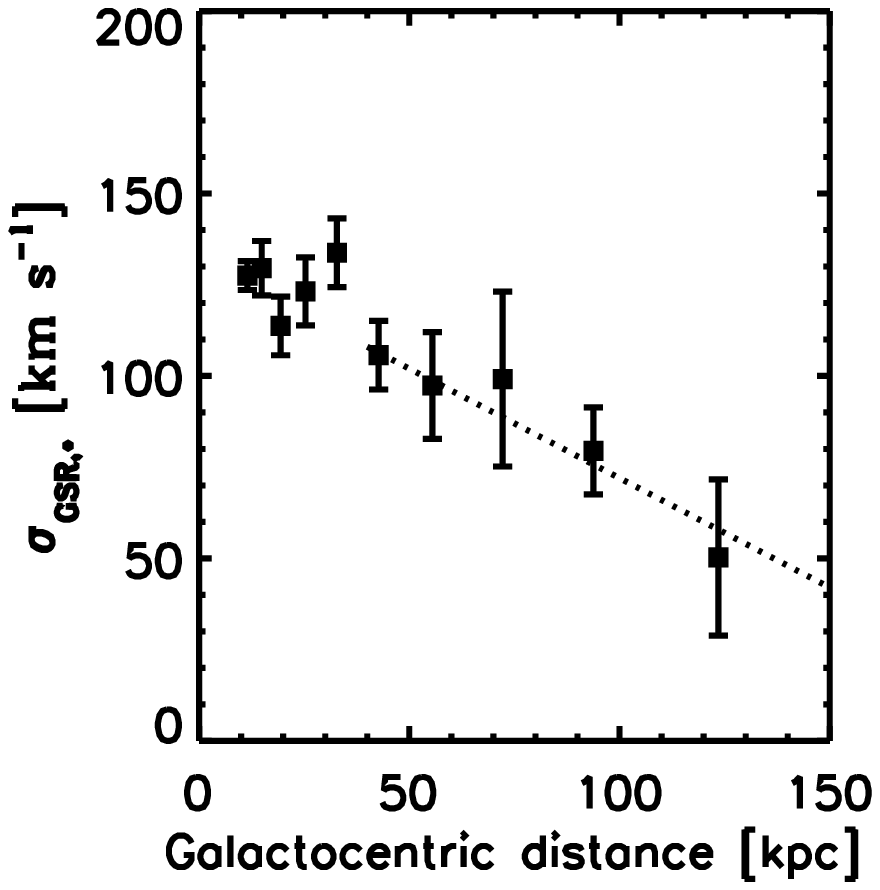}
\includegraphics[width=0.24\textwidth,clip]{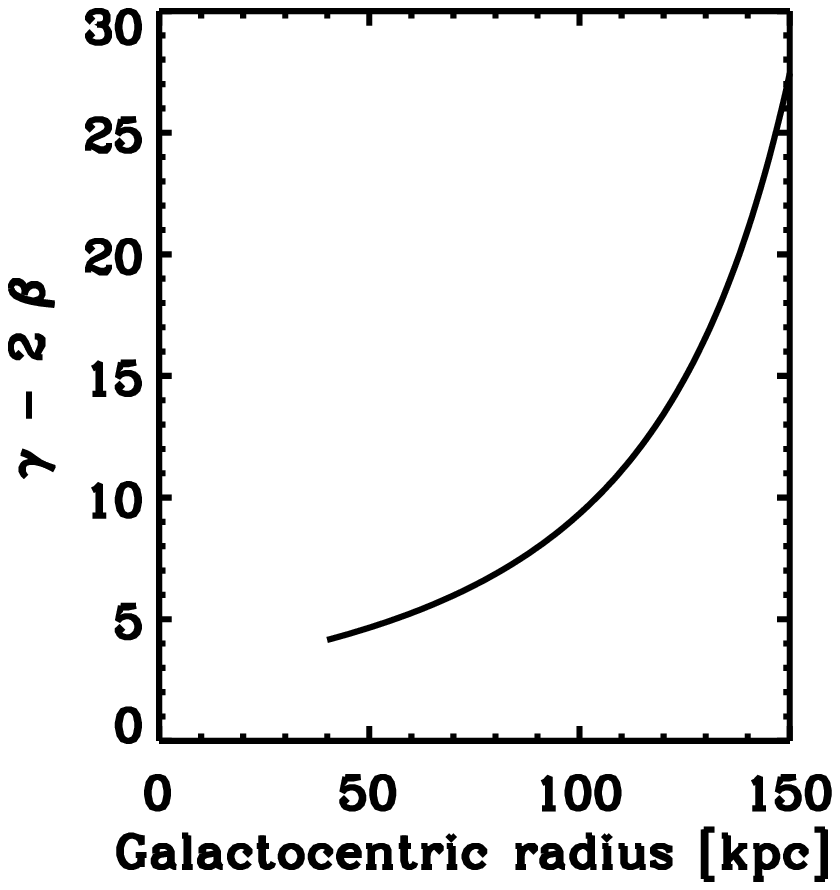}
\includegraphics[width=0.24\textwidth,clip]{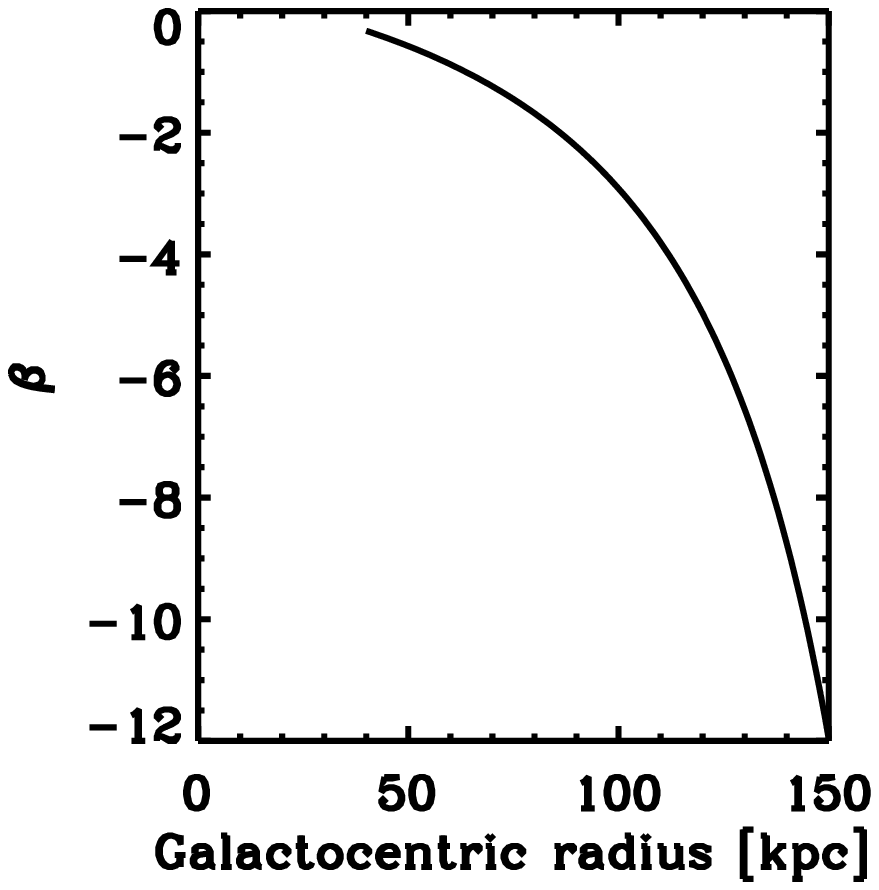}
\includegraphics[width=0.24\textwidth,clip]{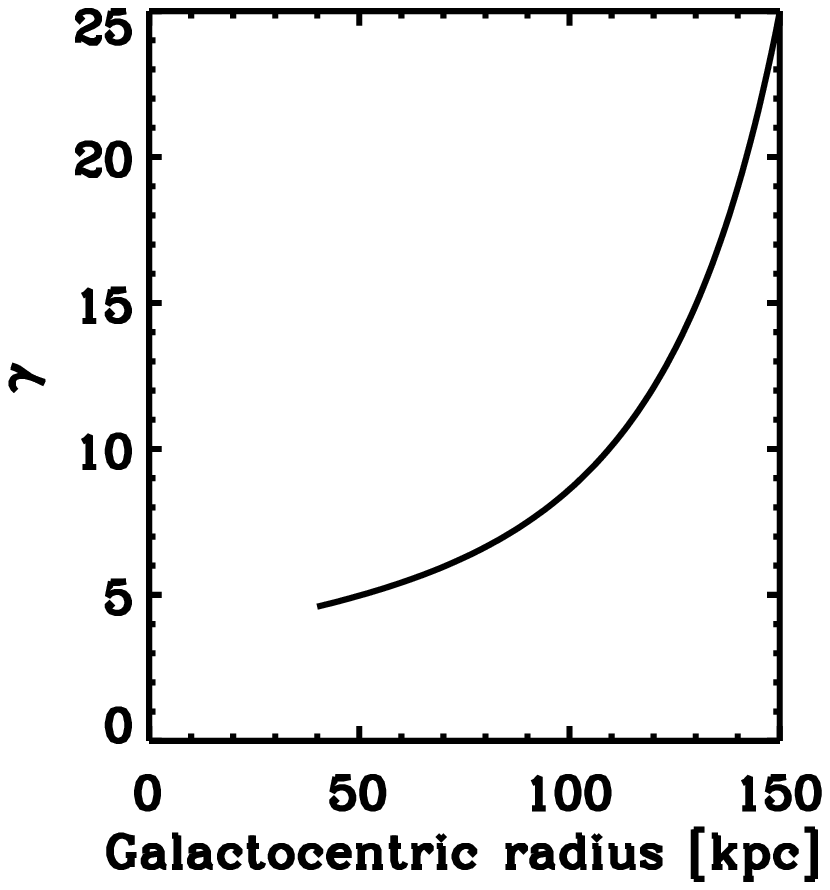}
\caption{Left panel: Observed Galactocentric radial velocity dispersion
(squares with errorbars); the dotted line is a straight line fit for
$r>$ 40 kpc. Second panel: relation $\beta$ and $\gamma$ should satisfy
to result in the same $\sigma{\rm_{GSR,*}}$. Third panel: variation of
$\beta$ with radius fixing $\gamma=$3.5. Right panel: variation of
$\gamma$ with radius fixing $\beta$ to the $\beta$-tg$_{\rm SN}$
model.}
\label{fig:app}
\end{figure*}

\section*{Appendix B}

Equation (2) shows that the radial velocity dispersion profile 
depends on the circular velocity given by the dominant mass component 
(i.e. the dark matter halo), 
the velocity anisotropy parameter $\beta$ and 
the power $\gamma$ of the density profile of the tracer population. 
For constant $\beta$ and $\gamma$, we can rewrite Eq.(2) as 
\begin{equation}
{\sigma^2}_{r,*}(r) = {1\over{r^{2\beta - \gamma}}}{\int_r}^{\infty}
V_{\rm c}^2(r') \, {r'}^{\,2\beta-\gamma - 1} \, dr'.
\end{equation}
In our work we assumed $\gamma=$ 3.5 at all Galactocentric distances,
but the above equation shows also that for a fixed mass distribution
(i.e. fixed circular velocity), models with the same value for
$2\beta-\gamma$ give rise to the same radial velocity dispersion
profile. In this Section we explore how $\beta$ or $\gamma$ have to
vary together in order to reproduce the observed Galactocentric radial
velocity dispersion.

In this analysis we restrict ourselves to Galactocentric distances
larger than 40 kpc, where: the value of $\gamma$ starts to become more
uncertain, the observed Galactocentric radial velocity dispersion
declines and the correction factor between the Galactocentric and the
true radial velocity dispersions is negligible.

At these distances the Galactocentric radial velocity dispersion
profile is well represented by a straight line, $\sigma_{\rm GSR,fit}=
a\,r + b$, with $a= -$0.6 and $b=$ 132 (Fig.~\ref{fig:app}, left).
We assume that the circular velocity for the dark matter halo is
constant and we fix it to $V_{\rm c}(r)= V_{\rm c}=$ 220 \kms. By
solving the Eq.(2) we obtain
\begin{equation}
\sigma_{r,*}^2= \frac{V_{\rm c}^2}{\gamma-2\beta}
\end{equation}
For all the values of $\beta$ and $\gamma$ that satify this relation
(at every $r$) the predicted radial velocity dispersion curve will be
the same. By imposing $\sigma_{r,*}^2=\sigma_{\rm GSR,fit}^2$ in
Eq.(19), it follows
\begin{equation}
\gamma-2\beta=  \frac{{V_{\rm c}}^2}{\sigma_{\rm GSR,fit}^2}
\end{equation}
Figure \ref{fig:app} (second panel) shows the above relation for the
assumed model.  The third panel in Fig.~\ref{fig:app} shows how
$\beta$ has to vary with the Galactocentric distance for this model if
we fix $\gamma=$3.5, whilst the panel on the right shows how $\gamma$
has to change if we use the $\beta$-tg$_{\rm SN}$ model for $\beta$.
Clearly for this model the values the $\gamma$ should assume in order
to reproduce the data are unrealistic.

The same kind of relation between $\beta$ and $\gamma$ can be derived
for different circular velocities in the regime where they can be
approximated by power-laws.

\label{lastpage}
\end{document}